\documentclass[journal=jpccck,manuscript=article]{achemso}

\usepackage{graphicx}
\usepackage{amsmath}

\title{Substitutional tin acceptor states in black phosphorus}

\author{Mark Wentink}
\affiliation{Department of Electronic and Electrical Engineering, University College London, WC1E 7JE, London, UK}
\altaffiliation{London Centre for Nanotechnology, University College London, WC1H 0AH, London, UK}

\author{Julian Gaberle}
\affiliation{Department of Physics and Astronomy, University College London, WC1E 6BT, London, UK}

\author{Martik Aghajanian}
\affiliation{Departments of Materials and Physics and the Thomas Young Centre for Theory and Simulation of Materials, Imperial College London, SW7 2AZ, London, UK}

\author{Arash~A.~Mostofi}
\affiliation{Departments of Materials and Physics and the Thomas Young Centre for Theory and Simulation of Materials, Imperial College London, SW7 2AZ, London, UK}

\author{Neil~J.~Curson}
\affiliation{Department of Electronic and Electrical Engineering, University College London, WC1E 7JE, London, UK}
\altaffiliation{London Centre for Nanotechnology, University College London, WC1H 0AH, London, UK}

\author{Johannes~Lischner}
\affiliation{Departments of Materials and Physics and the Thomas Young Centre for Theory and Simulation of Materials, Imperial College London, SW7 2AZ, London, UK}

\author{Steven R. Schofield}
\affiliation{Department of Physics and Astronomy, University College London, WC1E 6BT, London, UK}
\altaffiliation{London Centre for Nanotechnology, University College London, WC1H 0AH, London, UK}

\author{Alexander L. Shluger}
\affiliation{Department of Physics and Astronomy, University College London, WC1E 6BT, London, UK}

\author{Anthony J. Kenyon}
\affiliation{Department of Electronic and Electrical Engineering, University College London, WC1E 7JE, London, UK}
\email{a.kenyon@ucl.ac.uk}

\begin{document}

\date{\today}

\begin{abstract}
Nominally-pure black phosphorus (BP) is commonly found to be a p-type semiconductor, suggesting the ubiquitious presence of impurity species or intrinsic, charged defects.  Moreover, scanning tunnelling microscopy (STM) images of black phosphorus reveal the presence of long-ranged double-lobed defect features superimposed onto the surface atomic lattice.  We show that both the p-type doping of BP and the defect features observed in STM images can be attributed to substitutional tin impurities.  We show that black phosphorus samples produced through two common synthesis pathways contain tin impurities, and we demonstrate that the ground state of substitutional tin impurities is negatively charged for a wide range of Fermi level positions within the BP bandgap.  The localised negative charge of the tin impurities induces hydrogenic states in the bandgap and it is the 2p level that sits at the valence band edge that gives rise to the double-lobed features observed in STM images.  

\end{abstract}

%%%%%%% SECTION %%%%%%%%
\section{Introduction}

Black phosphorus (BP) is a two-dimensional (2D) van der Waals layered material and is considered a promising candidate for future (opto)electronic and quantum-electronic devices~\cite{Kim2019,Jia2019}.  BP holds a unique position within the family of 2D van der Waals layered materials, being the only mono-elemental 2D material besides graphite~\cite{Novoselov2004}.  Electrical and optical characterisation of nominally pure samples of BP uniformly find the material to be electronically p-type, dating back to the earliest measurements in the 1960s~\cite{Warschauer1963,Akahama1983,Castellanos-Gomez2015,Kiraly2017,Qiu2017}.  This suggests the ubiquitous presence of charged impurities or structural defects in BP; however, the identity of these impurities/defects, and therefore the source of the p-type doping in BP, remains unexplained.   In addition to causing shifts in the Fermi level, atomic point defects can greatly affect the fundamental properties of crystalline materials; for example they can act as nucleation centers for exotic phenomena such as charge density waves or superconducting phases~\cite{Li_2012,McChesney2010}.  Thus, determining the atomic structure of the intrinsic defects/impurities in BP is of critical importance for both the fundamental understanding of BP, and for nanoelectronic applications. 

Black phosphorus is comprised of hexagonally bonded sheets of phosphorus atoms, similar to the hexagonally bonded sheets of carbon atoms that form graphene.  However, because phosphorus has one extra valence electron than carbon, the phosphorus atoms of BP are $sp^3$ hybridised with three bonds to nearest neighbours and an electron lone pair, rather than $sp^2$ hybridised with a single electron (per carbon atom) in a $\pi$ orbital as is the case for graphene sheets.  This results in the individual phosphorene sheets of BP forming a buckled configuration, as indicated schematically in Fig.~\ref{fig:overview}a. The directions parallel and perpendicular to these rows are conventionally referred to as the zig-zag and armchair directions, respectively.  In addition to this strong structural anisotropy, BP also exhibits both in-plane and out-of-plane electronic anisotropy, with the in-plane anisotropy characterised by strong changes in electron effective mass and mobility between the zig-zag and armchair directions~\cite{Xia2014,Rodin2014}.

The presence of point defects in nominally intrinsic BP has been confirmed by atomic-resolution scanning tunnelling microscopy (STM) measuremts made by multiple groups~\cite{Kiraly2017,Qiu2017,Zhang2009}.  The defects appear as long-ranged anisotropic protrusions superimposed on the atomic lattice, and in the literature they have largely been attributed to charged single-phosphorus vacancies~\cite{Kiraly2017,Qiu2017,Riffle2018}. However, this assignment is in disagreement with density functional theory (DFT) calculations that demonstrate very small diffusion barriers for monovacancies in BP~\cite{Guo2015,Cai2016,Gaberle2018,Hu2015}.  These low barriers mean that monovacancies can be expected to diffuse much faster than STM scanning times at room temperature, and remain mobile even at 77~K.  However, no evidence of diffusion of the defects observed by STM have been reported~\cite{Kiraly2017,Qiu2017,Riffle2018}, consistent also with our observations here. DFT calculations also show that monovacancies are unstable with respect to the formation of divacancies.~\cite{Gaberle2018} Since divacancies are charge neutral, they are not expected to produce the long-ranged features observed in STM images, nor can they explain the p-type doping observed in BP.  Hence vacancy defects do not provide an explanation for the experimentally measured p-type electronic behaviour of BP, or the appearance of the defects observed in STM experiments. 

Here, we use a combination of  atomic-resolution STM, scanning tunnelling spectroscopy (STS), X-ray photoelectron spectroscopy (XPS), DFT, and tight-binding simulations, to investigate the origin of p-type doping in BP and the long-range defect features observed in STM images of BP.  We find that both the p-type doping and the defects seen in STM images can be attributed to substitutional tin impurities in the lattice, which are present due to the use of tin-based catalysts in BP growth, and that produce negatively-charged acceptor states in BP.  The characteristic double-lobed appearance of the defects in STM images results mainly from imaging a 2p orbital that exists just inside the bandgap at the valence band edge, with a lesser contribution also from a 1s orbital that lies deep within the bandgap.  

%%%%%%% SECTION %%%%%%%%
\section{Methods}

\subsection{Experimental methods}
Scanning Tunnelling microscopy and spectroscopy (STM/STS) measurements were performed using a commercial Scienta Omicron system operating at 77~K with a base pressure below 10\textsuperscript{-10}~mbar. Black phosphorus flakes were purchased from SmartElements and 2DSemiconductors. Samples with large single crystal areas were selected for STM measurement and mounted onto stainless steel sample holder plates using Epotek H21D conductive epoxy.  Samples were outgassed in vacuum at 150$^\circ$C overnight, before being cleaved in-situ to expose an atomically-clean surface. STM tips were chemically etched from tungsten wire and had their surface oxide removed by electron bombardment in the STM vacuum chamber. STS was performed by measuring the current as a function of the sample voltage with a constant tip-sample separation. X-ray photoemission spectroscopy (XPS) measurements were performed on a Thermo K-alpha system operating at a base pressure of 10$^{-9}$~mbar. Samples were cleaved ex-situ before rapid transfer into the loadlock and pumping down to 10\textsuperscript{-6} mbar. Cleaved samples were exposed to ambient conditions for less than one minute. 

\subsection{Density functional theory calculations}
Density functional theory (DFT) was used to calculate the electronic structure of substitutional tin defects in monolayer and multilayer BP. The calculations were performed using the CP2K code~\cite{Vandevondele2005}, which employs a mixed Gaussian and plane wave basis-set (GPW). The DZ\textunderscore MOLOPT\textunderscore GTH basis set was used together with Goedecker-Teter-Hutter (GTH) pseudopotentials~\cite{VandeVondele2007}. The plane wave cutoff was converged at 400 Ry, SCF convergence was set to $10^{-6}$ a.u. and residual forces on atoms were smaller than 0.01 eV/Å. Since GGA functionals predict a metallic behavior for BP, the PBE0-TC-LRC hybrid functional was used with a cutoff radius of 2~\AA~and 10\% Hartree-Fock (HF) exchange~\cite{functional}. In order to reduce the computational cost of the hybrid functional calculations, the auxiliary density matrix method (ADMM) was used, which uses a reduced basis set for the HF exchange calculation and thus enables the study of supercells up to 1500 atoms~\cite{ADMM} 

The optimized unit cell vectors were found to be $a = 3.4$~\AA, $b = 4.5$~\AA, and $c = 10.9$~\AA\ which is in good agreement with experiment and other DFT studies. The bandgap for bulk BP was calculated to be 0.5 eV, which increased to 1.35 eV for a monolayer. The surface properties of a thick sample used in this work were calculated using a 1296 atom cell ($9\times9\times4$ supercell) and the monolayer consisted of 324 atoms ($9\times9$) up to 1296 atoms ($18\times18$) cells with 20~\AA~vacuum gap. In order to estimate the band offset with electrodes, the ionization potential of the slab was calculated with respect to the vacuum level. For a monolayer, the VBM is at $-5.5$~eV, but increases to -5.2~eV for a bilayer and $-5.1$~eV for a four layer slab.

To calculate the defect formation energies and charge transition levels, the chemical potentials of P and Sn atoms in the gas phase were used.  All calculations included the potential alignment; however, we did not include charge corrections when calculating the charged states of Sn impurity but rather relied on the fact that these states are strongly delocalized and Sn atoms are positioned far away from each other due to large supercells. For a monolayer, we could achieve the formation energy convergence with the cell size.

\subsection{Tight binding method}
Tight-binding calculations were carried out to study the electronic structure of BP with a charged Sn substitutional defect. The BP surface is modelled as a monolayer that sits on a substrate with the dielectric constant of BP. For the BP monolayer, tight-binding parameters of Rudenko and Katsnelson\cite{Rudenko2014} were used. As the band gap of the monolayer is significantly larger than what is measured on the BP surface, we first diagonalize the Hamiltonian for the defect-free system and then apply a "scissor" shift to correct the band gap. Next, the defect potential is expressed in the basis of scissor-corrected eigenstates and the Hamiltonian is diagonalized again. To model the charged defect, the screened Coulomb potential induced by a point charge located at the position of the Sn nucleus (in the relaxed DFT geometry) is added as an on-site term to the Hamiltonian. For the screened potential, we employ an anisotropic Keldysh model of the form $V_\text{K}(x,y) = \frac{Z\pi e^2}{2r_0}[H_0(\epsilon_\text{env}\frac{r(x,y)}{r_0})-Y_0(\frac{\epsilon_\text{env}r(x,y)}{r_0})]$, where Z=-1 denotes the defect charge, $r_0=22.25$~\AA\cite{Rodin2014} and $\epsilon_\text{env}=10.5$ denotes the environment dielectric constant (determined by averaging the dielectric constants of bulk BP and of vacuum). Moreover, $r(x,y)=\sqrt{(\lambda_xx)^2+(\lambda_yy)^2 + d^2}$ describes the anisotropy of the screening with $\lambda_x=0.42$, $\lambda_y=1.2$ and $d=0.35$~\AA. The values of $d$ (which denotes the height of the Sn nucleus above the BP sheet), $\lambda_x$ and $\lambda_y$ were fitted to obtain good agreement with the measured tunneling spectrum of the defect. To obtain converged results for a single charged defect, we employ a $60 \times 60$ supercell and use a $3\times3$ k-point sampling of the corresponding first Brillouin zone. 

%%%%%%% SECTION %%%%%%%%
\section{Results}

\subsection{STM imaging and chemical composition of BP samples}

\begin{figure}[t!]
\centering
\includegraphics[width=12cm]{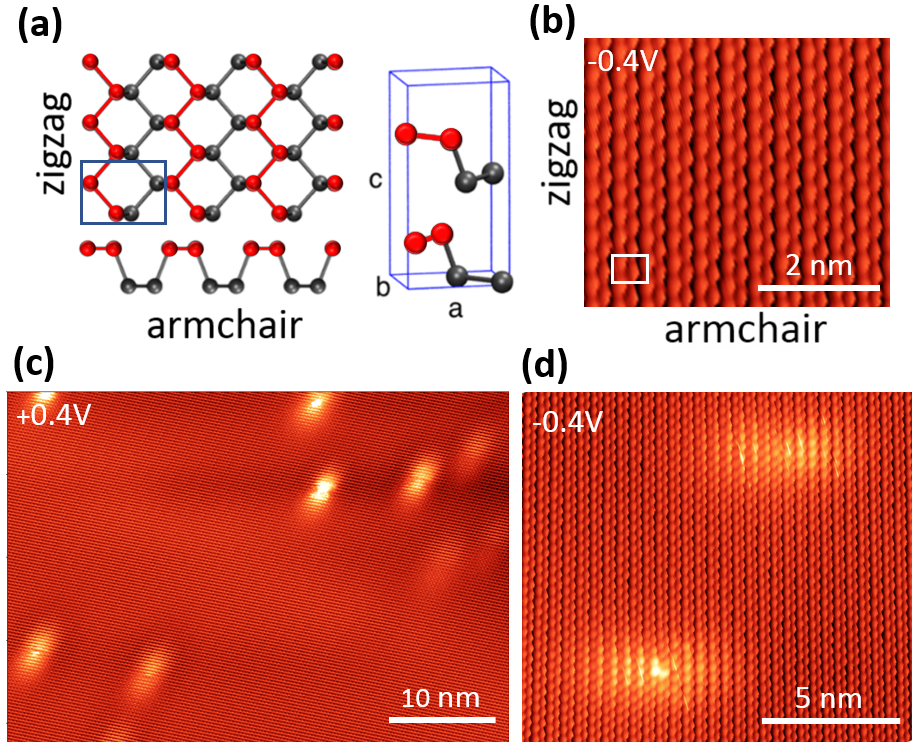}
\caption[overview]{(a) Structural model of a single phosphorene sheet shown in top and side views.  The phosphorus atoms at the top and bottom of the sheet are colored red and black, respectively; typical STM images resolve only the top (red colored) atoms.  Also shown is a single unit cell of BP, which shows two adjacent phosphorene layers in the AB stacking configuration.  (b) Atomic-resolution STM image of a clean and defect free area of a BP surface (image parameters: $-0.4$~V, 40~pA).  (c) Large area STM image showing multiple ($\sim10$) defects that appear as long range elongated protrusions ($0.4$~V, 40~pA). (d) Atomic-resolution image showing a pair of defects ($-0.4$~V, 40~pA). The zig-zag rows are aligned with the vertical axis of the image in panels (b) and (d), but run diagonally in panel (c).}
\label{fig:overview}
\end{figure}

Bulk BP crystals were cleaved under ultrahigh vacuum and imaged with STM at 77~K (see Methods).  An atomic-resolution STM image showing the zig-zag rows of the topmost atoms of a BP crystal is shown in Fig.~\ref{fig:overview}b. Fourier analysis of these images (not shown) yields lattice constants of $3.5\pm0.1$~\AA~and $4.2\pm0.1$~\AA~in the zigzag and armchair directions, respectively, in agreement with our density-functional theory calculations and previous studies~\cite{Kiraly2017,Qiu2017,Gaberle2018}. A larger area image where 10 defects can be observed is shown in Fig.~\ref{fig:overview}c.  These defects are typical of those in the literature~\cite{Kiraly2017,Qiu2017,Zhang2009} and present double-lobed protrusions that can extend as far as 10~nm in their long-axis direction (aligned to the armchair direction of the crystal).  A higher resolution image showing a surface region where two defects can be seen is shown in Fig.~\ref{fig:overview}d. 

\begin{figure}[h]
	\centering
	\includegraphics[width=16cm]{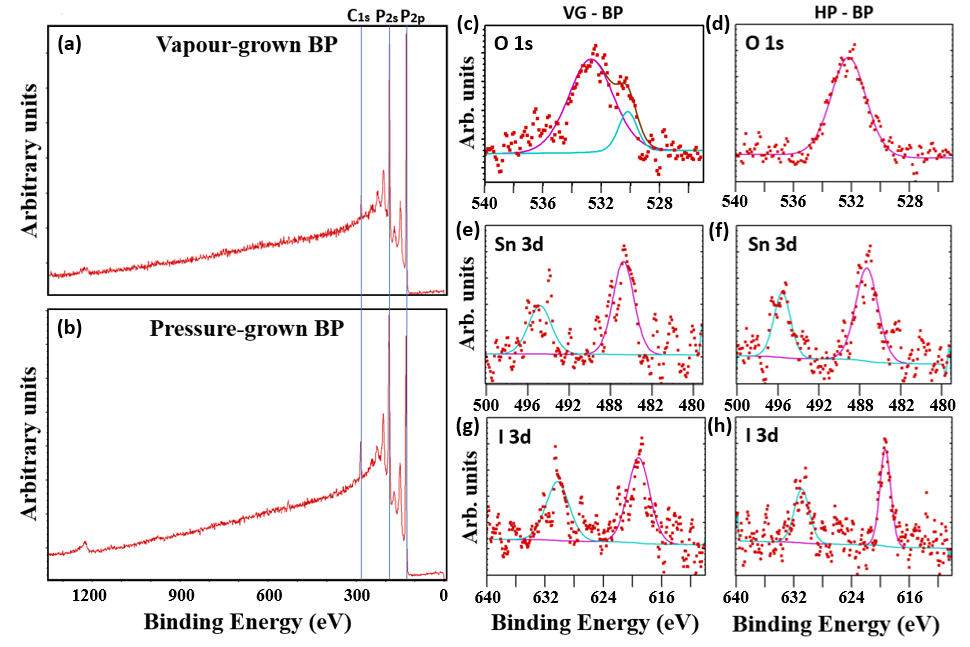}
	\caption[chemical]{XPS analysis of vapour-grown (VG) and high-pressure (HP) grown black phosphorus. Left: survey spectra resolving P and C peaks. The broad peak at 1230eV is the C KLL Auger peak. Right: high-resolution spectra of present impurities. O 1s peaks are observed at 532.5 and 530.2eV, the Sn 3d doublet at 487.0 and 495.2eV, and the I 3d doublet at 619.2 and 630.1eV}
	\label{fig:chemical}
\end{figure}

We use commercial BP samples obtained from two different suppliers: samples obtained from Smart-elements GmbH were fabricated using a vapour transport method~\cite{Lange2007,Kopf2014}, while samples obtained from 2D Semiconductors Inc. were fabricated via a high pressure conversion of red phosphorus~\cite{Bridgman1914,Shirotani1982}.  We refer to these samples as ``vapour grown'' and ``high pressure'' BP, respectively.  To test for the presence of chemical impurities, we have analysed freshly cleaved samples of both sample types using XPS.  Figures~\ref{fig:chemical}a,b show XPS survey scans for both sample types.  All the major peaks in this scan belong to phosphorus with the only other feature being trace carbon contamination, likely due to low-level hydrocarbon background in the high-vacuum chamber used for the measurement.  Additional contamination peaks (O, Sn, and I) were found using higher resolution and sensitivity XPS measurements, and we present these results in Figs.~\ref{fig:chemical}c--h.  The presence of an oxygen peak at 532.5~eV for both samples (Figs.~\ref{fig:chemical}c,d) can be attributed to physisorbed water, since the samples were cleaved in ambient conditions before being immediately introduced to the vacuum chamber.  The vapor grown sample (Fig.~\ref{fig:chemical}c) exhibits a smaller component at 530.2~eV, which may indicate some trace oxidation of the sample~\cite{Luo2016}; this difference between the two samples can be attributed to the significantly smaller crystal sizes of vapour grown BP, producing a higher density of crystal edges and rendering the samples more vulnerable to oxidisation~\cite{Kuntz2017}. Tin and iodine peaks were also found in both sample types (Figs.~\ref{fig:chemical}e--h).  The observation of these elements in our samples is in agreement with a previous report~\cite{Mayorga-Martinez2018} and can be attributed to the use of tin and iodine as catalysts during synthesis.  Similar high resolution scans in other regions (e.g., Fe, Cu, Ti, Zn) did not produce any measurable peaks.  It is notable that for both of our sample types the tin and iodine peaks are close to the measurement limit of our instrument.

\subsection{DFT calculations of tin and iodine impurities in BP}

The observation of tin and iodine in our samples suggests the possibility that one or both of these impurity species is responsible for the p-type doping of BP and/or the observed long-range defect features seen in STM images.  Recent DFT calculations examined a wide range of impurities in BP in adsorbed, intercalated, and substitutional lattice positions~\cite{Gaberle2018}.  Iodine was found to produce the least stable structures of all the elements considered.  The most stable iodine structure was an adsorbed neutral iodine atom, which did not produce any electronic states in the BP bandgap~\cite{Gaberle2018}. The absence of a localised charge or bandgap state rules out iodine being responsible for the features observed in STM images. In contrast, the substitutional tin impurity was found to be the most stable of all impurity defects considered~\cite{Gaberle2018}.  To better understand the defect behaviour as it is moved from the surface into the bulk, Sn impurities were calculated in various depths in a four layer BP slab. The slab consisted of 1296 atoms ($9\times9$ surface unit cell) and a single substitutional Sn was calculated in the upper row of the top layer, in the bottom row of the top layer and similarly in the second layer. As the Sn atom is moved into the slab the formation energy decreases slightly. The positively charged Sn in the top row of the top layer has the formation energy of -3.50 eV, while that in the second layer is slightly reduced to -3.56 eV.

\begin{figure}[h]
	\centering
	\includegraphics[width=12cm]{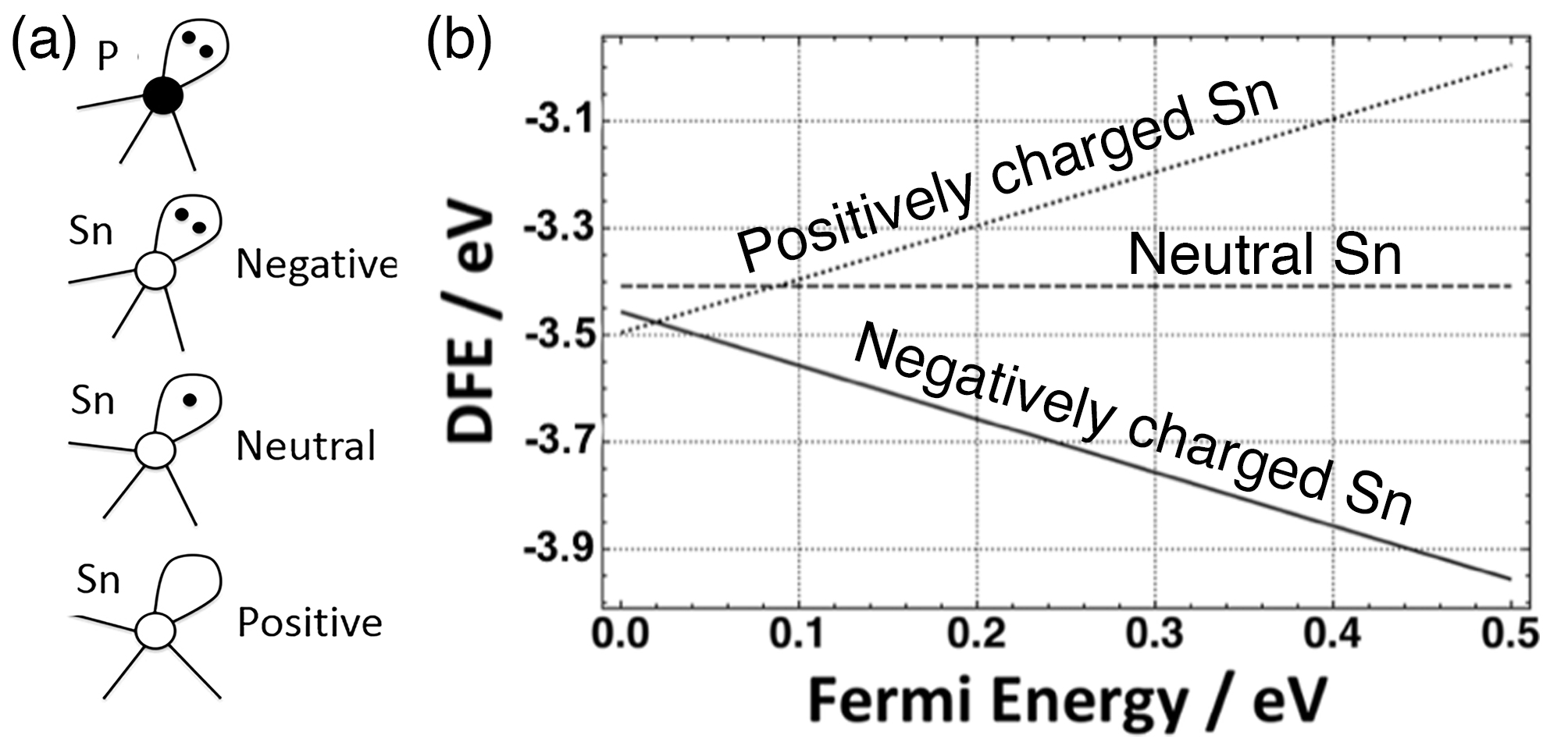}
	\caption[dft]{DFT modelling of substitutional Sn in black phosphorus. (a): Substitutional Sn adopts the same configuration as P, replacing the lone pair with a dangling bond that can hold 0, 1 or 2 electrons. (b): Formation energy of charged Sn defects for varying Fermi level. 0.0~eV corresponds to the valence band maximum.}
	\label{fig:dft}
\end{figure}

To investigate whether substitutional tin impurities might be the cause of the p-type doping in BP and the features observed in STM images, it is necessary to determine the relative stability of the three possible charge states of the defect (negative, neutral, and positive) in the presence of an electron reservoir with a chemical potential equal to the Fermi level of the crystal. To achieve this, we calculate the formation energies of the positive and negative charge states of the substitutional tin impurity at the surface of the four layer slab by adding or subtracting a single electron from the computational cell~\cite{Schofield2013a}.  The resulting defect states are drawn schematically in Fig.~\ref{fig:dft}a.  Each phosphorus atom in BP is three-fold coordinated with an electron lone pair.  Tin has one less valence electron than phosphorus, such that the neutral state of tin in BP features a singly-occupied dangling bond orbital.  The negative charge state of the substitutional defect has one additional electron, forming a lone pair, while the positive charge state has one less electron than the neutral case, leaving an unoccupied dangling bond orbital on the tin atom.  The relative stability of the three charge states as a function of the Fermi level, referenced to the valence band maximum, is ploted in Fig. \ref{fig:dft}(b). From this charge transition level plot, we see that the negative charge state of the substitutional tin impurity is energetically favourable for Fermi level values across most of the bandgap; the positively charged state is favourable only when the Fermi level lies within 0.02~eV of the valence band edge, and the neutral state is never favoured.  This demonstrates that an isolated substitutional tin impurity in BP will naturally adopt a negatively charged configuration by capturing an electron from the BP conduction band.  As the number of substitutional tin impurities increases, this will naturally lead to a shift of the Fermi level toward the valence band in an analogous manner to acceptor states in cubic semiconductors. The above results suggest that the p-type doping commonly observed in BP samples can be attributed to substitutional tin impurities.  

\subsection{Tunneling spectroscopy of surface defects}

Next, we consider whether substitutional tin impurities can explain the long-ranged defect features commonly observed in STM images of BP.  In cubic semiconductors, such as silicon and gallium arsenide, long-ranged features that appear superimposed on the surface atomic lattice have been attributed to presence of bulk substitutional dopant impurities located within several nanometres of the surface~\cite{Liu2001a,Yakunin2004a,Sinthiptharakoon2014a}.  Early reports found isotropic circular protrusions and depressions that were attributed to the effect of a localised charge-induced band bending, i.e., imaging the screened Coulomb potential of the dopant atom~\cite{Liu2001a}.  Later reports discovered that in some cases it is also possible to find highly anisotropic protrusions, and these features have been attributed to directly imaging the probability distributions of the hydrogenic states of substitutional dopants~\cite{Yakunin2004a,Sinthiptharakoon2014a}.  The highly anisotropic, double-lobed protrusions that are seen in BP (Figs.~\ref{fig:overview}b-d) are thus suggestive of imaging anisotropic hydrogenic states induced by a substitutional dopant.  

\begin{figure}[h]
	\centering
	\includegraphics[width=0.8\linewidth]{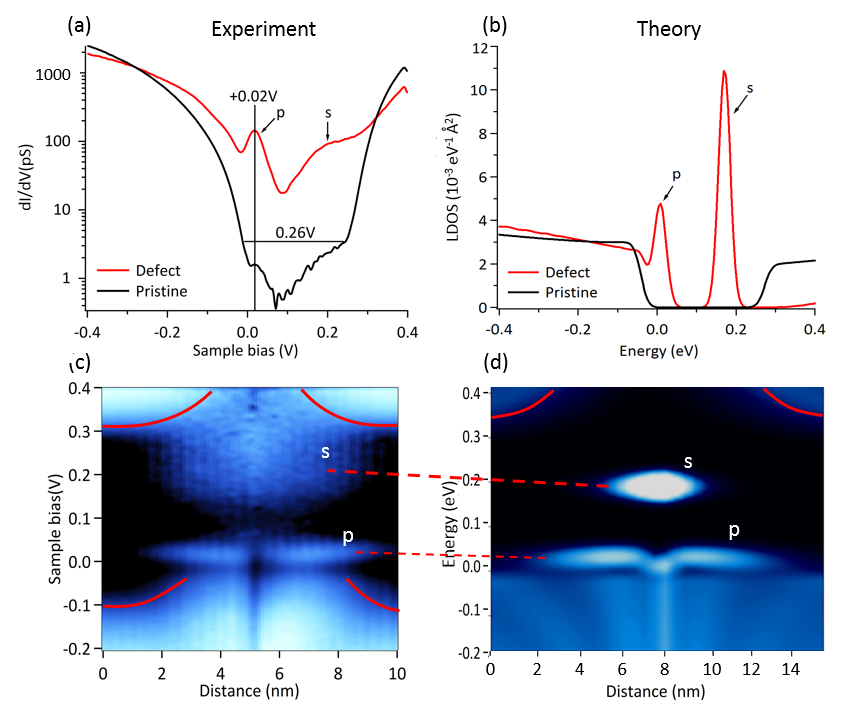}
	\caption[sts]{Comparison of experimental spectroscopy and tight-binding calculations. (a) and (b): experimental and simulated dI/dV curves on the bare surface (black) and in the vicinity of a double-lobed defect (red). (c) and (d): Experimental and simulated spatially resolved STS along a line through a double-lobed defect, where the colour scale corresponds to conductance. }
	\label{fig:sts}
\end{figure}

To determine whether this is the case, we have performed tunnelling spectroscopy measurements on the defects.  Figure~\ref{fig:sts}a shows tunnelling spectroscopy measurements recorded over a defect-free area of the surface (black trace) and over a double-lobed feature (red trace).  From the measurement over the defect free region we are able to extract a bandgap of $0.26\pm0.2$~eV, in agreement with previous reports~\cite{Prytz2010}, and we confirm that our sample is p-type~\cite{Akahama1983,Castellanos-Gomez2015,Kiraly2017,Qiu2017}. Examining the STS measurement over the defect, we notice a bandgap state peak at $\sim0.02$~eV, just above the valence band maximum.  We also notice that the defect measurement shows a shoulder at $\sim0.12$~eV, suggestive of a second peak deep within the bandgap.  These features are consistent with previously published spectra of defects in BP~\cite{Kiraly2017}.

Figure~\ref{fig:sts}c shows tunnelling spectrum as a function of position (horizontal axis) along the centre line of a single defect.  The left and right regions (near 0~nm and 10~nm) show the spectrum acquired over the clean surface and are in agreement with the point spectroscopy measurements in Fig.~\ref{fig:sts}a (black trace).  The central region of the measurement, i.e., between $\sim 2$ and 8~nm, shows the spatial variation of the tunnelling spectroscopy along the length of the defect.  In particular, the bandgap state that was seen at $\sim0.02$~eV in the point spectrum (Fig.~\ref{fig:sts}a) is resolved into a double-lobed feature with a node in the centre extending $\sim 8$~nm across the surface; this is the same spatial variation as the features observed in the topographic images, demonstrating that this bandgap state is the origin of the double-lobed features observed in STM images.  The broad shoulder at $\sim0.12$~eV appears as a diffuse region of enhanced conductivity, notably appearing as a continuous region with a maximum intensity near the centre rather than a node as in the case for the 0.02~eV feature.  We do also see some upward band bending of the band edges on either side of the defect, consistent with the defect carrying a negative point charge. 

\subsection{Calculations of hydrogenic states of tin at the BP surface}

In order to explain the physical characteristics of the defect state and determine whether it can be attributed to substitutional tin, it is necessary to perform simulations with unit cells large enough to model hydrogenic wave functions, and to capture the effects of charge-induced band bending.  Such unit cells consist of many thousands of atoms, and are not currently accessible using DFT.  Therefore, we have modelled the hydrogenic states produced by a substitutional tin impurity using tight-binding calculations.  In these calculations, the defect is modelled as a negative point charge at the position of the tin nucleus (see Methods), which represents the negative charge that is localised on the lone pair orbital of the substutional tin impurity as predicted by our DFT calculations.  We plot the simulated local density of states (LDOS) resulting from the tight-binding calculation in Fig.~\ref{fig:sts}b; the red and black traces show the calculation with and without the negative point charge, respectively, and these can be compared directly to the corresponding experimentally measured tunnelling point spectra in Fig.~\ref{fig:sts}a.

The simulated LDOS of the defect-free surface (black trace, Fig.~\ref{fig:sts}b) reproduces the bandgap seen in the experimental data (black trace, Fig.~\ref{fig:sts}a). The simulated LDOS of the defect (red trace, Fig.~\ref{fig:sts}b) shows a bandgap state near the valence band maximum and a second peak near 0.18~eV, in good agreement with the experimental data (red trace, Fig.~\ref{fig:sts}a). The spatial variation of the LDOS across a defect is shown in Fig.~\ref{fig:sts}d and can be compared to the experimental measurement in Fig.~\ref{fig:sts}c.  Here we see that the double-lobed state just inside the bandgap at the valence band edge is reproduced in our tight bonding model, as is the upward band bending of the conduction band edge.  The tight binding simulation also shows a nodeless state close to mid-gap, that corresponds well to the nodeless state observed in the experimental data.  

The gap states in the tight-binding simulations are hydrogenic states induced by the localised negative point charge placed in the lattice to represent the negatively-charged substitutional tin impurity.  This negative charge creates a partially screened Coulomb potential and induces hydrogenic defect states.  Analysis of the simulation data shows that the two bandgap states in Fig.~\ref{fig:sts}b can be attributed to the 1s hydrogenic ground state, and the 2p$_x$ hydrogenic excited state (where $x$ corresponds to the heavy hole, i.e., armchair, direction), while the 2s and 2p$_y$ states are shifted into the BP valence band. It is the anisotropic electronic structure of BP that results in the reordering of the state energies compared with the normal spherical hydrogenic model.  

%%%%%%% SECTION %%%%%%%%
\section{Conclusion}

We have investigated the origin of p-type doping in nominally-intrinsic BP, and whether this doping can be attributed to the long-ranged defects commonly observed in STM images of BP. Using the combination of XPS, atomic-resolution STM/STS, and DFT, we demonstrated that tin impurities are present in BP samples, and that substitutional tin impurities are both highly stable and act as acceptor states. Using tight-binding calculations, we have demonstrated that substitutional tin impurities induce hydrogenic defect states in BP, and that the 2p state of these defects is the main source of the long-range features observed in STM images of BP.  These observations provide compelling evidence that substitutional tin impurities are responsible for the p-type doping in BP, and are the origin of the long-ranged defect features observed in STM images of BP. 

\section*{Acknowledgements}
MW and MA were supported through studentships in the EPSRC Centre for Doctoral Training in Advanced Characterisation of Materials (EP/L015277/1) and EPSRC Centre for Doctoral Training in Theory and Simulation of Materials (EP/L015579/1), respectively. AAM, JL, ALS, MA, JG, and AJK acknowledge the support of the Thomas Young Centre through grant TYC-101.  ALS and AJK acknowledge funding provided by EPSRC (EP/P013503/1), and Leverhulme Trust (RPG-2016-135). SRS acknowledges funding provided by EPSRC (EP/L002140/1).  JG acknowledges the use of the ARCHER high-performance computing facilities via the membership in the U.K.s HPC Materials Chemistry Consortium, funded by EPSRC (EP/L000202) and the computer resources of the UK Materials and Molecular Modelling Hub, which is partially funded by EPSRC (EP/P020194).

%%%%%%% BIBLIOGRAPHY %%%%%%%%
%\bibliography{manuscript}

\providecommand{\latin}[1]{#1}
\makeatletter
\providecommand{\doi}
  {\begingroup\let\do\@makeother\dospecials
  \catcode`\{=1 \catcode`\}=2 \doi@aux}
\providecommand{\doi@aux}[1]{\endgroup\texttt{#1}}
\makeatother
\providecommand*\mcitethebibliography{\thebibliography}
\csname @ifundefined\endcsname{endmcitethebibliography}
  {\let\endmcitethebibliography\endthebibliography}{}

\end{document}